# Review of 5G capabilities for smart manufacturing

Joachim Sachs
Ericsson Research
Stockholm, Sweden

Krister Landernäs
ABB Corporate Research
Västerås, Sweden

*Abstract*—5G has been defined to address new use cases beyond consumer-focused mobile broadband services. In particular industrial use cases, for example in smart manufacturing, have been addressed in the 5G standardization, so that 5G can support industrial IoT services and wireless industrial networking. To this end, 5G needs to integrate with the industrial network based on Ethernet and TSN. 5G can create new opportunities for smart manufacturing by enabling flexibility and increasing the automation in the production. This paper provides an overview of the use cases and requirements for smart manufacturing that can be addressed with 5G and which are validated in three industrial 5G trial systems. The capabilities of 5G are described for providing non-public networks that support industrial LAN services based on Ethernet and TSN. The 5G radio access network functionality is described and discussed for practical deployments. The paper provides an overview of the current state-of-the-art of using 5G for smart manufacturing.

*Keywords—5G System, non-public networks, TSN, industrial IoT, industry 4.0, edge cloud*

## I. Introduction

Two trends that are driving the development of smart manufacturing are 1) more flexible production, to meet the need of increased customization, and 2) more autonomous operations and monitoring to increase productivity and improve quality [1]. The digital transformation of industry fueled by Industry 4.0 and the development of 5G cellular technology, is in many ways a response to these trends. Work on defining 5G requirements for industry use cases has been done both in 3GPP standardization [2] and 5G-ACIA [4][5][6]. This work is expected to continue also in the future as the 5G technology evolves to support even more use cases.

## II. Use cases and requirements

### A. Selecting a Template (Heading 2)

In general, Industry 4.0 use cases are quite diverse ranging from control applications that are very demanding in terms of end to end (E2E) quality of service (QoS) to monitoring applications with more relaxed requirements. In the 5G-SMART project, seven use cases for smart manufacturing are described [9] including e.g., mobile robotics, wireless sensors for process monitoring and integration with industrial LAN. An overview of different use cases and their respective classification is given in Table I.

TABLE I. 5G-SMART USE CASE CLASSIFICATION

| 5G-SMART Use cases (according to 3GPP TR 22.804) | | Factory automation | Process automation | HMIs and Production IT | Logistics and warehousing |
|---|---|---|---|---|---|
| UC1 | 5G-Connected Robot and Remotely Supported Collaboration | X | | | X |
| UC2 | Machine Vision Assisted Real-time Human-Robot Interaction over 5G | X | | X | X |
| UC3 | 5G-Aided Visualization of the Factory Floor | X | | X | |
| UC4 | 5G for Wireless Acoustic Workpiece Monitoring | X | X | | |
| UC5 | 5G Versatile Multi-Sensor Platform for Digital Twin | X | X | | |
| UC6 | Cloud-based Mobile Robotics | X | | | X |
| UC7 | TSN/Industrial LAN over 5G | X | | | |

Work on defining and classifying different industry use cases have been done by 5G-ACIA [4]. Also, in this work, the diverse requirements in smart manufacturing are apparent as seen in Table II. The wide range of use cases introduces a challenge to provide communication services tailored to different applications while safeguarding critical communication.

A major factor for consideration when introducing new technology in factories is the large investment in *Operational Technology* (OT) equipment done by end customers. These systems are based on different industrial technologies and local network solutions. New network architecture concepts need to co-exist and seamlessly integrate with the existing solutions.

TABLE II. USE CASES AS DEFINED BY 5G-ACIA [4]

| | Use case | Category [a] |
|---|---|---|
| 1. | Connectivity for the factory floor | Hard RT |
| 2. | Seamless integration of wired and wireless components for motion control | Hard RT |
| 3. | Local control-to-control communication | Hard RT |
| 4. | Remote control-to-control communication | Soft RT |
| 5. | Mobile robots and AGVs | Soft RT |
| 6. | Closed-loop control for process automation | Soft RT |
| 7. | Remote monitoring for process automation | Non-RT |

[a.] Hard real-time (RT) / Soft RT / Non-RT refer to highly critical / moderately critical / not critical latencies, see [4]

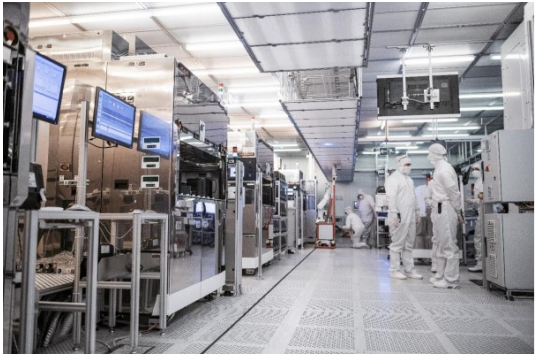

Fig. 1. Inside a clean room of a semiconductor factory (see [10][11])

Radio coverage in a factory environment can be challenging. Even in large factories assembly lines and machinery are usually located very dense making line-of-sight communication hard to achieve (see Fig. 1). Furthermore, factory floors are usually environments characterized as rich scattering with various tools and machines, which contribute to shadowing effects and are in different ways interacting with the radio signals. Moreover, the environment is typically quite dynamic as production is ongoing with people and robots moving around performing their tasks. Careful radio planning should therefore be done before deploying a 5G system in a factory.

In addition to technical enablers in 5G such as low latency communication and network slicing, there are different architectural options that will accelerate the adoption of 5G into smart manufacturing ecosystem. These include 5G support for Ethernet-based industrial networks and edge computing that enables data processing close to the industrial equipment. To this end the following requirements on 5G can be defined:

- For industrial applications 5G needs to provide a dedicated network service to the industry customer with guaranteed availability and a high level of data security and privacy.

- A dedicated 5G network needs to integrate with existing industrial networks. The primary wired networking technology in industrial networks is Ethernet, which is complemented with several special-purpose communication technologies such as wired fieldbuses / real-time Ethernet variants, and wireless sensor networks. This heterogeneous communication technology landscape is evolving towards a consolidation of the network design to few openly standardized communication technologies. With the addition of TSN for deterministic low latency communication, Ethernet/TSN will become the central multi-service wired communication technology in industrial deployments. Similarly, 5G is expected to provide the wireless connectivity for the range of industrial services, and to integrate seamlessly with the wired network infrastructure.

- A converged communication infrastructure opens opportunities to bring novel capabilities of cloud computing to the industrial systems. This includes cost-effective resources for scalable compute and storage, which is facilitator to introducing novel capabilities, such as data analytics, AI/ML, machine vision, digital twins. These capabilities should be enabled throughout the shopfloor and manufacturing system.

- Manufacturing comprises many different services from sensor collection to real-time control. 5G needs to provide the capabilities to address all relevant use cases and provide sufficient wireless capacity. In particular the addition of *ultra-reliable and low latency communication* (URLLC) and *time-sensitive communication* (TSC) is a game changing capability of 5G that enables wireless connectivity also for deterministic time-critical industrial services, such as closed-loop control [2][4][5][6][7].

III. DEDICATED 5G NETWORKS FOR INDUSTRIAL SYSTEMS

A. 5G-Non Public Networks

When deploying a 5G network for an industrial manufacturing plant, the 5G network needs to be integrated with the industrial equipment and installations, see Fig. 2. Generally, a dedicated 5G network is required that is customized to the needs and services of the factory. To this end, 3GPP has defined 5G *non-public networks* (NPN) that are reserved for private connectivity of a defined group of 5G devices to connect to an industrial network [3][8][14][21]. An NPN can be realized in two variants: as *standalone NPN* (SNPN) or as *public-network integrated NPN* (PNI-NPN).

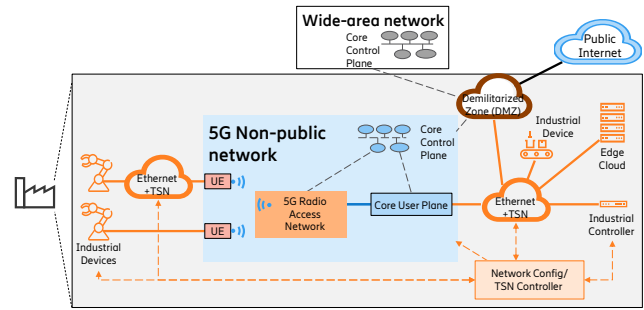

Fig. 2. Industrial 5G network

In an SNPN all network services are realized dedicated for the industrial user; dedicated network nodes are deployed for the SNPN service and are typically located on-premises of the industrial site. The SNPN has its own network identifier; only devices being part of this SNPN are authorized to register to the network and obtain connectivity. In contrast, for a PNI-NPN some functionality is shared between the NPN and a *public network* (PN). The PNI-NPN uses a public network identifier, to which both public devices and NPN devices can generally connect. To separate the private PNI-NPN service from public network services, the PNI-NPN is e.g., provided by means of a dedicated network slice. Access to the PNI-NPN is restricted to a defined group of authorized devices. The network slice of the PNI-NPN can be isolated from the network functions and services that are used by public network devices. In order to provide guaranteed network capacity and availability, network resources can be reserved and allocated to the PNI-NPN; also dedicated network functions can be provided to the PNI-NPN. Typically, a PNI-NPN for factory comprises a dedicated network deployment on-premise of the industrial site. This includes for example, the radio network that provides the radio coverage and capacity, but typically also the core network *user plane function* (UPF) as gateway to the industrial network for user plane traffic. In that way, no private user data leaves the industry premises.

The radio cells on the industrial site can be reserved for exclusive use by the PNI-NPN users. Access restrictions can be applied on tracking area level or on radio cell level. For example, certain cells can be configured for a *closed access group* (CAG) so that only PNI-NPN devices are allowed to access those radio cells. A benefit of a PNI-NPN is that the industrial network service can be extended beyond the industrial premises. For example, also logistics in-between two industrial sites, or inbound/outbound logistics to/from a factory can be integrated into the industrial system. The PNI-NPN service is then provided via the antenna masts deployed for the public network, e.g., around the industrial premises. The service area of the PNI-NPN can be controlled per tracking area, or even on cell level via CAGs if needed.

Different operational models are possible for NPNs. A mobile network operator can provide both PNI-NPN or SNPN solutions to an industrial network, which can comprise the installation and operation of the 5G network. For an SNPN it is also possible that an industrial entity builds and operates its own 5G NPN. In this case, the industrial partly needs to have spectrum rights for the operation of a 5G network, as discussed later in section IV. Generally, multiple operation models are possible for NPNs, which can involve the industrial party, a mobile network operator and further 3rd party system integrators. Operational models are discussed in [14].

*B. LAN services via 5G*

An industrial network is providing LAN services to the connected devices and equipment. When a 5G NPN is deployed in the factory, it needs to integrate with the Ethernet LAN services on the shopfloor, as shown in Fig. 2. On the field level and within specific network segments, it is typically complemented with dedicated wired communication technologies based on fieldbus technologies (see IEC 61784-1) or some real-time Ethernet variants (see IEC 61784-2), which often deviate from the IEEE 802.1 Ethernet standard.

5G supports the transport of standard IEEE 802.3 Ethernet traffic since Release 15. The 5G system (5GS), comprising the 5G network and the 5G *user equipment* (UE), appears as a virtual Ethernet bridge and is prepared for integrating with existing *network management systems* (NMS), typically using network management protocols such as *Simple Network Management Protocol* (SNMP) / *Management Information Base* as shown in Fig. 3. The 5GS bridge can in this way expose its capabilities including bridge and port management information (e.g. LLDP, VLAN), and the 5GS bridge can obtain configuration information from the NMS in particular related to IEEE 802.1Q. A 5GS deployment in an integrated Ethernet network architecture can have multiple 5G virtual Ethernet bridges. Additionally, 5GS also support classic Ethernet forwarding mechanisms such as flooding and *Medium Access Control* (MAC) learning.

*Time-sensitive networking* (TSN) is a set of open IEEE standards that introduces capabilities of reliable and deterministic low-latency networking to the IEEE 802-LANs. TSN is over time expected to replace fieldbus and real-time Ethernet variants, leading to a common converged TSN-capable Ethernet LAN infrastructure for both normal Ethernet traffic together with time-critical deterministic TSN traffic.

3GPP has introduced support for integration with wired TSN in Release 16 with further additions in Release 17. That is, TSN real-time capabilities over standard Ethernet have been added to 5G, and the 5GS is modelled as a TSN capable Ethernet bridge, so that 5G can be integrated with TSN to address the relevant use cases for industrial automation ([7][13][14][15][16]). The 5GS behaves towards the Ethernet network as a set of virtual TSN-capable Ethernet bridges, which provide Ethernet ports on the mobile device side (with a *device-side TSN translator* (DS-TT) at the UE) and the network side (with a *network side TSN translator* (NW-TT) at the UPF) as shown in Fig. 4. Ethernet and TSN communication are possible in-between any of those Ethernet ports over the 5G network. 5G provides a control plane function (TSN AF) which interacts with the central TSN control plane function as defined in IEEE 802.1Qcc: the TSN *centralized network configuration* (CNC). TSN functionality may be applied in only in parts of an industrial network, e.g., when a segment of an Ethernet network is upgraded with TSN support.

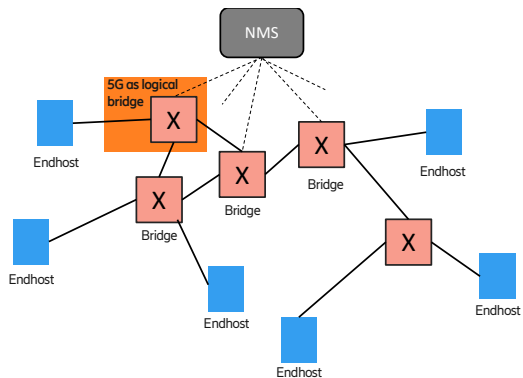

Fig. 3. 5G integration with Ethernet LAN

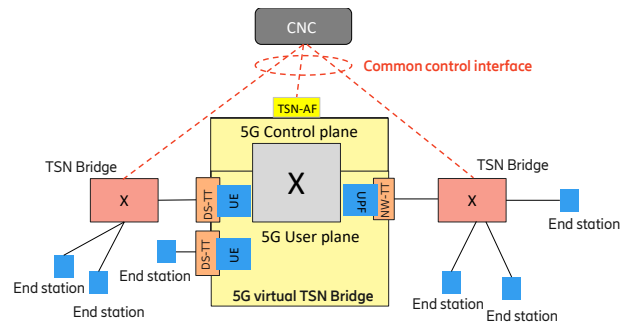

Fig. 4. 5G as virtual TSN bridge

Transport of Ethernet traffic is provided within the 5GS in *Ethernet PDU sessions*. 5G applies a QoS framework by differentiation traffic in differing *QoS flows*, which can be associated with different priority level, packet delay budget and tolerable packet error rates. Ethernet applies prioritization by different traffic streams by means of *priority code points* according to IEEE 802.1Q. Via the TSN AF the 5G bridge properties and TSN functions are configured. Ethernet/TSN traffic streams are mapped to 5G-specific QoS flows based on the Ethernet priority code point. According to the traffic requirements, 5G can configure service-specific treatment for different QoS flows. For example, for time-critical traffic, QoS flows for ultra-reliable transmission with ultra-low latency can be configured according to the URLLC capabilities that have been specified for 5G. In addition, TSN features such as per-stream filtering and policing (IEEE 802.1Qci), e.g. to protect against misbehaving traffic, and time scheduling (IEEE 802.1Qbv) for handling of critical traffic classes can be supported by 5G in the TSN translator functions.

### C. Time synchronization

Advanced manufacturing applications can benefit from 5G capabilities that go beyond data communication, like time synchronization. Time synchronization among industrial equipment allows to perform coordinated tasks or observe a sequence-of-events in industrial measurements. In addition, some traffic handling configurations for Ethernet/TSN require time-based configuration of traffic handling across multiple TSN bridges, which need to be synchronized to a common reference time.

5G provides the capability to transfer a time reference from a reference grandmaster (GM) clock to an end station across the 5GS while maintaining the time error introduced by the 5GS below 900 ns. To enable this, the 5GS provides a common reference to both the UPF (for example, via a transport network) and to the UE (via the radio interface), as shown in Fig. 5. When time synchronization is transmitted via the *(generalized) precision time protocol*, (g)PTP, between any grandmaster clock to an end station, the 5GS can determine the residence time of the PTP message within the 5GS and correct the time value in the PTP message accordingly [3][13][15][16]. In a similar way, multiple PTP time domains can be supported by the 5GS in order to transfer multiple time references (GMs) to differing end stations and applications.

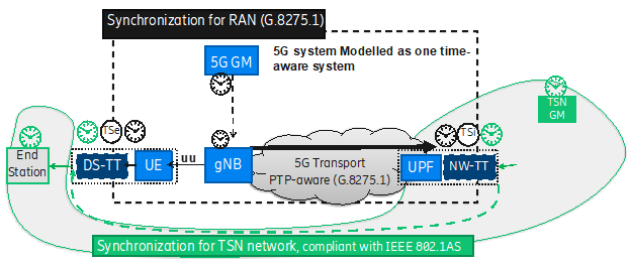

Fig. 5. Time synchronization over 5G

### D. Edge cloud

Cloud computing provides novel opportunities for industrial automation. Traditionally, application logic for industrial automation is provided in dedicated hardware. Often this dedicated compute hardware is integrated into the industrial equipment, like the robot or the machine. Moving such functionality into a cloud execution environment provides access to efficient and scalable compute and storage resources. It allows to profit from the dynamic technical developments in the compute area; novel functionality can be provided as part of regular maintenance of the central compute infrastructure. Due to e.g., data privacy policies and stringent performance requirements, the cloud computing paradigm in industrial deployments has a strong emphasis on edge computing, which means that the computing infrastructure is close to the application and typically located on the industry premises, as shown in Fig. 2.

Moving application functionality from industrial equipment to an edge cloud brings several benefits. The equipment gets lighter, uses less energy and can be simpler maintained. But also new functionality can be flexibly introduced: collaborative tasks can be provided by coordination of the application functions within the same cloud platform, allowing that e.g. a common map is jointly built up and shared between a set of mobile robots or automated guided vehicles [9]. Similarly, if a video system is installed on the shopfloor the information extracted by machine vision can be exploited for different tasks, as steering robot action according to visual guidance, safety assessment of human workers on a shopfloor, or remote monitoring/surveillance.

The adoption of cloud computing brings on the other hand new requirements on the communication infrastructure to connect the industrial equipment, sensors, actuators, and devices for human-machine interactions with the edge compute infrastructure. A fully connected network infrastructure is needed that provides ubiquitous network access to the edge cloud on the entire shopfloor, in contrast to the hierarchically segmented industrial networks of today. This network infrastructure needs to provide very high availability and also ensure that deterministic latency critical service performance can be guaranteed, to e.g. critical control loops. The broad adoption of Ethernet/TSN and 5G provides the wired and wireless infrastructure for this: a converged multi-service communication infrastructure with support for time-sensitive deterministic communication.

## IV. 5G Radio Network Design

### A. Spectrum options

5G radio networks can operate in different spectrum bands. Licensed spectrum bands that have been defined today cover the range of frequency-division duplex (FDD) of 450 MHz - 2.7 GHz and time-division duplex (TDD) in the ranges 1.8 - 4.9 GHz and 24 - 48 GHz. In addition, 5G can operate in unlicensed spectrum bands at 5 and 6 GHz. The bands that are available differ etween countries. 5G spectrum with highest relevance for local industrial deployments is licensed spectrum in midband (ca. 1.8 - 5 GHz) and highband (ca. 24 - 48 GHz) [10]. In contrast to unlicensed spectrum, licensed bands enable network deployment with very high availability that support time-critical

URLLC services, as the access to the spectrum resources can be fully controlled. These spectrum bands are further well suited for local deployments and provide wide bandwidth with good capacity.

Generally, 5G spectrum is licensed to mobile network operators (MNOs). Those have multiple spectrum bands available that can be used for NPN deployments. Several countries have recently also adopted local spectrum licenses, which enable an industrial entity to obtain a local spectrum license for its premises for non-public network services [17]. Local spectrum options in those countries can allow up to 100 MHz of TDD spectrum in midband, and more than 100 MHz of TDD spectrum in highband.

*B. 5G Ultra-reliable and low latency communication*

5G introduces mechanisms for ultra-reliable and low latency communication. This comprises features to enable faster transmissions, as well as functionality to increase the reliability of transmissions. The transmission reliability can be increased by novel robust transmission modes for data and control channels, the usage of multi-antenna transmissions or redundant transmissions of data over multiple radio links and carriers. If the delay bound permits, fast HARQ retransmissions can be used to increase reliability in a spectrally efficient way, but introduce a latency increase as shown in Table III. The latency of the transmission depends on the length of the radio slots. 5G introduces short slot structures and fast processing times, with the option of using even shorter mini-slots for time critical data. Further latency reductions are achieved by pre-allocating transmission resource to time-sensitive data flows. In TDD bands, the latency is also depending on the TDD pattern with which transmission opportunities are alternating between uplink (U) and downlink (D) transmission. More frequent alternations between uplink and downlink reduce the latency as shown in Table III. More information on URLLC can be found in [18][19][20][21].

TABLE III. 5G RAN LATENCIES IN DOWNLINK (DL) AND UPLINK (UL) FOR DIFFERENT TDD PATTERNS AND NUMBER OF RETRANSMISSIONS (FROM [10])

| Latency in ms | DDDU 14 [b] | | DUDU 14 | | DUDU 7 | |
|---|---|---|---|---|---|---|
| | *DL* | *UL* | *DL* | *UL* | *DL* | *UL* |
| 1st Tx | 1.68 | 2.68 | 1.68 | 1.68 | 0.93 | 0.93 |
| 2nd Tx | 3.68 | 4.68 | 2.68 | 2.68 | 1.93 | 1.93 |
| 3rd Tx | 5.68 | 6.68 | 3.68 | 3.68 | 2.93 | 2.93 |
| 4th Tx | 7.68 | 8.68 | 4.68 | 4.68 | 3.93 | 3.93 |
| 5th Tx | 9.68 | 10.68 | 5.68 | 5.68 | 4.93 | 4.93 |
| 1st Tx | 1.68 | 2.68 | 1.68 | 1.68 | 0.93 | 0.93 |

[b]. "DDDU 14" corresponds to 3 DL slots, followed by 1 UL slot, with each slot being 14 OFDM symbols long. "3th Tx" corresponds to one failed transmission followed by 2 HARQ retransmissions

The capacity that a 5G radio network can provide depends not only on the traffic characteristics but also on the service requirements. Table IV shows the 5G radio capacity for a 5G network deployment with 100 MHz of midband TDD spectrum in a reference factory and with a 5G configuration optimized for URLLC. Periodic closed loop control traffic with small packet size is investigated. Different service requirements are considered regarding the latency bound for the data transmission and the reliability with which the timely data delivery is ensured. The capacity is defined as the traffic load, at which the service requirements can still be met for all devices throughout the factory shopfloor, i.e., with 100% service availability. More details on the evaluation are found in [10]. Note that the capacity increases for more relaxed service requirements. If the latency requirements of the service allow for retransmissions, HARQ retransmissions can be applied to increase spectrally efficiency tool for achieving the desired reliability level. For ultra-low latencies, other mechanisms need to be applied to achieve reliability by e.g., extra robust transmission modes and redundant transmissions, which reduce the spectral efficiency of the transmission. The total system capacity depends also strongly on the deployment of the 5G RAN in the factory, e.g. how many base stations are deployed. An extensive analysis of RAN performance can be found in [10].

TABLE IV. 5G RAN CAPACITY IN DOWNLINK AND UPLINK IN MB/S FOR A RAN DEPLOYMENT IN MIDBAND FOR DIFFERENT TDD PATTERNS AND SERVICE REQUIREMENTS ON LATENCY AND RELIABILITY (FROM [10])

| Latency | Reliability | DDDU 14 [b] | | DUDU 14 | | DUDU 7 | |
|---|---|---|---|---|---|---|---|
| | | *DL* | *UL* | *DL* | *UL* | *DL* | *UL* |
| 1 ms | 99.999% | - | - | - | - | 20 | 61 |
| | 99.9% | - | - | - | - | 99 | 110 |
| 2 ms | 99.999% | - | - | 22 | 60 | 130 | 145 |
| | 99.9% | - | - | 105 | 105 | 200 | 230 |
| 3 ms | 99.999% | 33 | 30 | 145 | 150 | 190 | 220 |
| | 99.9% | 160 | 54 | 220 | 230 | 265 | 310 |
| 5 ms | 99.999% | 220 | 74 | 265 | 280 | 265 | 310 |
| | 99.9% | 335 | 115 | 320 | 330 | 300 | 345 |

What can be seen in Table IV is that ultra-low latencies of 1 2 ms require latency-optimized TDD patterns, where uplink and downlink slots are frequently alternating. While 5G allows for flexible configuration of TDD patterns, coexistence with other networks needs to be considered. If two neighboring 5G networks apply different TDD configurations, cross-link interference situations can occur where an uplink transmission in one network interferes with a downlink transmission in the other network and vice-versa as shown in Fig. 6. Good isolation (e.g. by wall loss) or separation between neighboring networks is required in this case to avoid harmful interference. If the isolation is not sufficient, coordination mechanisms for interference reduction may be needed between the networks, or the networks need to apply a common synchronized TDD configuration. Synchronization of the TDD patterns in networks is common practice between mobile network operators for outdoor macro networks. For a 5G NPN deployment, the need for synchronized TDD configuration limits the flexibility of TDD pattern that can be applied. A deeper analysis of coexistence can be found in [10].

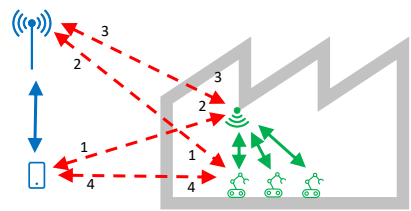

Fig. 6. Coexistence between a NPN and a neighboring network

## V. Industrial 5G Validation

In the 5G-SMART project, the implementation and validation of several trial use cases are being carried out at three trial sites located across Europe: in Kista (Sweden), Aachen (Germany) and Reutlingen (Germany). The three trial facilities bring different perspectives; the Kista trial site focuses on industrial robotics use cases (UC1-UC3 in Table I), the Aachen trial site focuses on integrating 5G into the networked adaptive production (UC4-UC5 in Table I), and the Reutlingen trial site validates 5G in a real operational factory environment (UC6-UC7 in Table I), as well as analyses channel propagation characteristics. The trials provide a platform to validate 5G capabilities for current and future manufacturing applications. Specifically, the trials will consist of 5G SNPN deployed on-premises of the industrial sites, which include an edge cloud in the factory to support various manufacturing applications.

## VI. Summary

5G networks provide the capabilities to address new use cases for smart manufacturing. They can be deployed as non-public network with integrated edge cloud at industrial sites. With well-designed radio network deployment good coverage and capacity can be provided to industrial use cases. This paper provides an introduction into industrial use case and a 5G network design to address those. Three 5G trial systems are deployed in factory shopfloors in the 5G-SMART project to validate the benefits of 5G for smart manufacturing.

## Acknowledgment

This work has been performed in the framework of the H2020 project 5G-SMART co-funded by the EU. The authors would like to acknowledge the contributions of their colleagues. This information reflects the consortium's view, but the consortium is not liable for any use that may be made of any of the information contained therein.